\documentclass[12pt,fleqn]{article}
\usepackage{multicol}
\usepackage{graphicx}

\begin{document}
\sf
\centerline{\tt LNGS/TH-01/03 \hfill \tt  Accepted in Nucl.Phys.B}
\vspace{7mm}

\centerline{\LARGE\sc Mirror model for sterile neutrinos}

\vspace{6mm}

\centerline{\large\sf
Veniamin Berezinsky, Mohan Narayan, Francesco Vissani}
\centerline{\em INFN, Laboratori Nazionali del Gran Sasso,
I-67010 Assergi (AQ), Italia}
\vspace{5mm}

\centerline{\large\sc Abstract}
\begin{quote}
\small
Sterile neutrinos are studied as subdominant contribution to solar
neutrino physics. The mirror-matter neutrinos are considered as
sterile neutrinos. We use the symmetric mirror model with
gravitational communication between mirror and visible sectors. This 
communication term provides mixing between visible and mirror
neutrinos with the basic scale $\mu = v_{\rm EW}^2/M_{\rm Pl}=
5\times 10^{-6}$~eV, where $v_{\rm EW}=174$~GeV is the vacuum
expectation value of the standard 
electroweak  group and $M_{\rm Pl}$ is the 
Planckian mass. It is demonstrated
that each mass eigenstate of active neutrinos splits into two states 
separated by small $\Delta m^2$. Unsuppressed oscillations 
between active and sterile neutrinos ($\nu_a \leftrightarrow \nu_s$)
occur only in transitions between each of these close pairs
(``windows''). These oscillations are characterized by very small 
$\Delta m^2$ and can suppress the flux and distort spectrum of 
$pp$-neutrinos in detectable way. The other observable
effect is anomalous seasonal variation of neutrino flux, which
appears in 
LMA solution. The considered subdominant neutrino
oscillations $\nu_a \leftrightarrow \nu_s$ can reveal itself as 
big effects in observations of supernova 
neutrinos and high energy (HE) neutrinos. 
In the case of HE neutrinos they can provide a
very large
diffuse flux of active neutrinos 
unconstrained by the e-m cascade upper limit. 
\end{quote}
\rm

\section{Introduction}
\subsection{Why light sterile neutrinos?}
A sterile neutrino, $\nu_s$, is a neutral, spin 1/2 particle which is a
singlet under SU(3)$\times$ SU(2)$\times$U(1) (SM model) group. 
In supersymmetry, grand unified theories (GUT) and 
superstring models such singlets appear naturally. 
It is more difficult however to have them light. Sterile 
neutrino always interacts, sometimes very weakly, 
with ordinary neutrinos. 
This interaction naturally results in small neutrino mixing.
It is more difficult to build a model with large mixing between
sterile and ordinary neutrinos. 

However, there are many models of
light sterile neutrinos, which effectively 
mix with the ordinary ones. It is natural to consider the
models in which ordinary and sterile neutrinos mix, 
even if this effect is weak. In our view, an investigation 
of sterile neutrinos should not necessarily start 
from this or that observational motivation; it should rather 
select a well definite theoretical framework 
and study its implications, with the hope 
that some of them are observable. 
This is the strategy of our work. 

It must be said that at present there are not many indications  
for effective oscillations between 
sterile and active neutrinos. Two flavor oscillation 
$\nu_e\rightarrow \nu_s$ is excluded as solution to the solar neutrino
problem, especially after SNO data \cite{SNO}. 
Oscillation $\nu_{\mu}\rightarrow \nu_s$ is 
disfavored as solution of atmospheric neutrino anomaly \cite{atm}. 
In recent works \cite{goz,fogli,strumia,maltoni},
stringent upper limits on the contribution of sterile 
neutrinos to the solar or atmospheric 
neutrino experiments are  obtained.
An indirect hint for a sterile neutrino
comes only from interpretation of the above-mentioned  
data when combined with those of  LSND \cite{LSND} (for the difficulties 
in such interpretation see \cite{Valle}, 
and \cite{hp} for further discussion). 

However, a small mixing of sterile
neutrinos with active ones and/or very small $\Delta m^2$ remain a 
viable possibility. This possibility 
can reveal itself in {\em subdominant 
processes}, or in some {\em new phenomena} such as neutrino 
oscillations at very large distances: neutrinos from 
supernovae and high energy (HE) neutrinos from the mirror matter.  

The models for sterile neutrinos include those based on supersymmetric 
theory, with or without R-parity violation e.g.\ \cite{JoSm,Valle2}, 
on the GUT and string models, e.g.\ component of 27-plet in 
E$_6$ model \cite{E6}
or SM singlet with additional U(1) charge 
\cite{La}, on the hidden sector, modulinos in 
supergravity \cite {BeSm},  and many other models. 

We believe that {\em any} theoretical 
model for sterile neutrinos should explain why the mass 
of a sterile neutrino is close to that of the ordinary neutrinos.
Actually, this is the reason why we select the mirror models 
\cite{FoVo,BeMo,BeVi}    for  sterile neutrinos:  
neutrino masses arise only from operators 
with dimensions 5 or larger, and therefore are suppressed 
by inverse power of large masses, exactly as for ordinary neutrinos.

\begin{figure*}[t]
  \begin{center}
  \mbox{\includegraphics[width=0.6\textwidth]{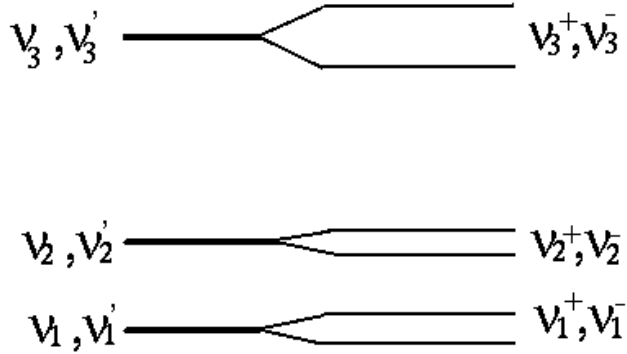}}
  \end{center}
  \caption{The double degeneracy between 
neutrino mass eigenstates of ordinary and mirror world 
($\nu_i$ and $\nu'_i$, respectively,
with $i=1,2,3$) is lifted when 
the small mixing (communication) terms are included.
The new mass eigenstates, 
denoted as $\nu_i^+$ and $\nu_i^-$, 
are maximal superpositions  
of $\nu_i$ and $\nu'_i$: $\nu_i^+=(\nu_i+\nu'_i)/\sqrt{2}$
and $\nu_i^-=(\nu'_i-\nu_i)/\sqrt{2}$.
}\label{fig:sp}
  \end{figure*}

 In this paper we discuss the origin of the mixing of ordinary 
and mirror neutrinos (`communication' term) 
and estimate its impact on oscillations. The most important effect 
is the splitting of the unperturbed mass eigenvalues of active neutrinos
to two close-set states (see Fig.~\ref{fig:sp}). The transition between 
these states results
in the oscillation to sterile neutrinos with long wavelength. 
We consider the applications of our model to solar,
supernova and high energy neutrinos. 

\subsection{Models with mirror matter}
Mirror matter was first suggested by Lee and Yang \cite{LeYa} in 
1956, who proposed that the transformation in the particle
space which corresponds to space inversion $\vec{x} \to -\vec{x}$ 
should not be the usual parity transformation $P$, but $P\times R$, 
where $R$ transforms a particle ({Lee and Yang considered the proton})
into a reflected state in the mirror particle space. This concept was further
developed by Salam \cite{Sa}, but 
in fact this idea was clearly formulated  
only later, in 1966,
by Kobzarev, Okun and Pomeranchuk \cite{KoOkPo}.
{ In this work it has been proposed that mirror and ordinary matter 
may communicate only gravitationally, and that the objects from mirror 
matter (stars and planets) can be present in the universe. Okun \cite{Okun}
considered also the communication due to new very weak long-range forces
and discussed this interaction for celestial bodies from mirror matter.} 
Since that time mirror matter has found interesting
phenomenological applications and development \cite{mirror,KoSeTu}. 
It has been boosted in 1980s by superstring theories with $E_8\times E_8'$ 
symmetry. The particle content and symmetry of interactions in each of
the $E_8$ groups are identical, and thus the mirror world has
naturally emerged. 

One can describe the motivation for the mirror sector in the following
way.
 
The Hilbert space of the particles is assumed to be a representation of 
the {\em extended} Poincar\'e group, i.e.\ one which includes the 
space coordinate inversion $\vec{x} \to -\vec{x}$. Since the coordinate 
operations, inversion and time shift, commute, the corresponding
operations in the particle space, $I_r$ and 
Hamiltonian $H$, must commute,
too. It implies that parity defined as 
eigenvalue of operator $I_r$ must be integral 
of motion for a closed system. For this it can be
suggested that $I_r=P\times R$, where 
$P$ is the usual operator of reflection 
and $R$ is the operator which transforms an ordinary 
particle to the mirror particle. Thus parity is conserved in the total 
Hilbert space of ordinary and mirror particles.  
The assumption of Landau \cite{Landau} was 
$R=C$, i.e.\ we can say that he suggested to use antiparticles 
as the mirror space, but then $CP$ must be conserved
which we know is not the case today. 

Mirror neutrinos as sterile neutrinos and various  applications of 
mirror matter have been intensively studied during the past
several years 
in the context of explanation of atmospheric and solar neutrino problems    
\cite{BeMo,FoVo,Foot}, cosmological problems, including inflation and 
nucleosynthesis \cite{BeDoMo,Berezh,FoVo97,Foot,BeVi}, dark matter and
galaxy formation \cite{BeDoMo,Berezh,MoTe,Berezh1}, extra dimensions 
\cite{Si} and high energy neutrinos \cite{BeVi}. 
    
Mirror matter scenarios have two basic versions. 
The {\em symmetric} version 
was suggested in the early works and more recently 
advocated in \cite{FoLeVo,FoVo}.
The Lagrangian which describes the particles and their interactions
in the visible and mirror sectors, ${\mathcal L}_{\rm vis}$ and 
${\mathcal L}_{\rm mirr}$, are perfectly symmetric and transforms into
each other when $\vec{x} \to -\vec{x}$, accompanying by all left
states transforming into right and vice versa: $\Psi_L \to \Psi_R'$
and $\Psi_R \to \Psi_L'$, where primes denote the mirror states.
The vacuum expectation values
(VEV's) of the Higgs fields are also identical in both sectors. 
Parity is conserved in the 
enlarged space of ordinary and mirror states. 
The masses and mixings of neutrinos in the {\em symmetric} mirror
model is studied by Foot and Volkas in \cite{FoVo}. The two sectors
(ordinary and mirror) communicate through the Higgs potential and
mixing of neutrinos. This is a phenomenological description, in
contrast to dynamical description of Refs.\cite{ABS,BeMo}, 
where the two sectors
interact gravitationally and neutrino mixing follows from this
interaction. Neutrino masses and mixings in each sector are 
induced by the usual see-saw mechanism, and mixing of neutrinos of
different sectors are postulated as e.g.\ $m {\nu}_L\nu_R'$, 
where mirrors neutrinos are denoted by primes. As demonstrated in 
\cite{FoVo} the most general mixing terms compatible with 
parity conservation results in maximal mixing of ordinary and sterile
neutrinos. 
 
The {\em asymmetric} version was suggested by Berezhiani and Mohapatra 
\cite{BeMo}. They assume that while all coupling constants in the two
sectors are identical, the VEV's are different and
break the parity. The ratio $\zeta=v'/v$ of electroweak VEV's 
($\langle H\rangle=v$ and $\langle H'\rangle =v'$)
gives thus the scaling factor for ratios of masses in the ordinary and
mirror worlds, such as masses of gauge bosons, leptons and quarks.
The basic communication between the 
two worlds is gravitational. It is taken in
the form of universal dimension 5 operators, 
suppressed by the Planckian mass 
$M_{\rm Pl}$. Operating inside each world and between them, these
terms give neutrino masses and mixings. However, to describe the
desired neutrino masses, the authors assume also additional 
communication through the singlet  superheavy fields, which results 
in the similar
dimension 5 operators suppressed 
by superheavy mass $\Lambda < M_{\rm Pl}$.  

A similar model--with asymmetric hidden sector--was studied in 
Ref.\cite{BeVi}. The communication of the two sectors is described by 
a dimension 5 operator with superheavy mass $\Lambda$ in the
denominator. The neutrinos in this model are found to be maximally
mixed and mass degenerate.
The neutrino masses and mixings are obtained
with help of dimension 5 operators with one scale $\Lambda$, 
with two different electroweak VEV's, $v$ and $v'$, in the 
visible and mirror sectors, respectively, and using VEV's of two 
SU(2) singlets $\langle \Phi \rangle =V$ and 
$\langle \Phi' \rangle =V'$.

\subsection{The nucleosynthesis constraints}
There are several dangers to be watched for in the models with 
mirror matter. They are connected with  
cosmological (big-bang) nucleosynthesis.

In mirror models the number of massless and light particles is doubled, 
and this case is excluded by cosmological nucleosynthesis if the
temperatures of mirror and ordinary matter is the same. 
A natural solution might be given by making the temperature of mirror
matter, $T'$, lower
than that of visible matter $T$. 
It is rather difficult to implement 
this in the symmetric models \cite{FoVo}. 
In asymmetric models \cite{BeMo} a natural 
solution is given by the different couplings of the inflaton to the 
visible and mirror particles \cite{BeDoMo}. The inflaton decays with 
different rates
to visible and mirror matter, producing thus the different
temperatures of the two sectors. In the symmetric models 
the different temperatures can be obtained in a
two-inflaton model \cite{BeVi}. In this model there are two inflatons,
$\varphi$ and $\varphi'$, with identical couplings to the visible and mirror 
matter, respectively.
The roll of the inflatons towards the minimum of the potential
is not synchronized, and the particles produced by the inflaton which reaches
the minimum earlier will be diluted by the 
inflation driven by the second inflaton.
By definition, the first inflaton is the mirror one. 

Having the temperature $T' < 0.5 T$, does not solve the problem of
cosmological nucleosynthesis completely. 
The number of additional effective
neutrino flavors  is limited by cosmological nucleosynthesis as 
$\Delta N_{\nu}< 0.2 -0.3$. Even if the initial density of mirror neutrinos 
is strongly suppressed, they might reappear 
again with the equilibrium density
due to oscillation of the visible neutrinos to sterile neutrinos; for 
a review and references see \cite{Do-rev}. 
The oscillations might bring the sterile
neutrinos in equilibrium with the active ones. (Indeed, while $\nu_a$'s
oscillate into $\nu_s$'s, the missing $\nu_a$'s are replenished again 
by thermal production.) 
In case of small mixing angles, when $\nu_a$ and $\nu_s$ can be
approximately considered as mass eigenstates with masses $m_a$ and $m_s$,
the non-resonant oscillation $\nu_a \to \nu_s$ occur when 
$\Delta m^2=m_s^2-m_a^2 >0$ . In this case the limit on 
$\Delta m^2$ allowed by nucleosynthesis is given by \cite{Do-rev}:
\begin{equation}
\Delta m^2\; \sin^42\theta \leq \xi\; 10^{-5}(\Delta N_{\nu})^2\ \mbox{eV}^2,
\label{nucl-vac}
\end{equation}
where $\theta$ is the vacuum mixing angle for $\nu_a-\nu_s$ mixing,
and $\xi=3.16$ for $\nu_a=\nu_e$, and  
$\xi=1.74$ for $\nu_a=\nu_{\mu/\tau}$. 
Using $\Delta N_{\nu}< 0.2 -0.3$ one obtains from Eq.~(\ref{nucl-vac})
the upper limit on $\Delta m^2$ for given mixing angle $\theta$.  

For large (or maximal) mixing, 
the upper bound on $\nu_a$ (where $a \neq e$) oscillations continues 
to roughly satisfy the scaling law previously given, 
whereas the bound on $\Delta m^2_{\nu_e\nu'}$ 
becomes much more stringent. 
In a number of calculations 
\cite{shi,enqv,md} bounds were obtained in the range 
\begin{equation}
\Delta m^2_{\nu_e\nu'}\leq  10^{-8}-10^{-9}~{\rm eV}^2,
\label{max-mix}
\end{equation}
where the effect of $\nu_a\to\nu_e$ oscillations are neglected.

The model for neutrino masses presented in this paper 
has oscillations into sterile neutrinos with small 
$\Delta m^2$, that satisfy the bounds cited above
(see Section \ref{m-spectrum}). However, we note that, 
{\em a priori},  these bounds are not unavoidable.
As was first remarked in Ref.\cite{FoVoL}, these limits  
become much weaker in presence of large lepton asymmetry 
$L=(n_{\nu}-n_{\bar{\nu}})/n_{\gamma}$. For example, in case of 
$\nu_e \to \nu_s$ oscillation the limit (\ref{nucl-vac}) is replaced by  
\begin{equation}
\Delta m^2/{\rm eV}^2 < 4\cdot 10^2|L_e|.
\label{nucl-L}
\end{equation}
The suppression of $\nu_a \to \nu_s$ oscillation 
is due to the matter effects, which appear because
the neutrino potential depends on lepton asymmetry. 
The scale for large and small lepton asymmetry is given by the baryon 
asymmetry $B \sim 10^{-10}$.
The lepton asymmetry needed for the above-mentioned effect must be larger
than $\sim 10^{-7}$ \cite{KiCh}. An application for atmospheric and
solar neutrino oscillations with $L \sim 10^{-5}$ is considered 
in \cite{FoVoN,Fo}.    
The lepton asymmetry can be generated by some unspecified mechanism,
but to be  larger than $B \sim 10^{-10}$, the lepton asymmetry must be 
generated after electroweak phase 
transition, otherwise it would be reprocessed
by the sphaleron mechanism into too large a baryon asymmetry. 
A mechanism of generation of the lepton asymmetry is neutrino
oscillation itself (see \cite{Do-rev} for the status and references).  
Lepton asymmetry is generated only in case of small neutrino mixing. 
At maximal mixing, for example, probability of $\nu_s \to \nu_a$ 
oscillation is large, and scattering of $\nu_a$ provides thus the
equilibrium of sterile and active neutrinos. Foot 
\cite{Fo} suggested the following model where suppression of oscillation 
is provided by self-generating lepton asymmetry. There are four neutrinos 
$\nu_{\tau}, \nu_{\mu}, \nu_e$ and $\nu_s$, among them  $\nu_{\tau}$ 
and $\nu_s$ 
have small mixing, while the other neutrinos are allowed to have the
large mixing. Large lepton asymmetry $L$ is produced by 
$\nu_{\tau} \to \nu_s$ oscillation, and $L$ suppresses the oscillations
of the other neutrinos. 
(The oscillation $\nu_a \to \nu'$, where $\nu'$ is the mirror neutrino, 
is suppressed more strongly than $\nu_a \to \nu_s$, because of
self-interaction of $\nu'$ \cite{FoVoN}.)
Thus, even in the case of large $\Delta m^2$, the $\nu_a \to \nu_s$ 
oscillation and the resulting nucleosynthesis restrictions are suppressed
in presence of large lepton asymmetry, either existing or
self-produced.

\section{The model}

\subsection{Origin of the mass terms}

Our model belongs to the more general framework of the symmetric models 
of Foot and Volkas \cite{FoVo}. Differently from these 
authors, we discuss the origin of the mass terms,
and focus on the models with  gravitational communication terms
between ordinary and mirror neutrinos.
 
\subsubsection{Gravitational communication terms}\label{communication}

As in the first classical works \cite{LeYa}-\cite{mirror} on
mirror matter we shall assume
that that mirror particles communicate with the visible ones only 
gravitationally. For the description of this interaction we shall
use dimension 5 operators \cite{d=5-,d=5}. 
For neutrinos these communication term, obtained from the
$SU(2)_L\times U(1)\times SU(2)'_R\times U(1)'$ scalar, 
reads:
\begin{equation}
{\mathcal L_{\rm comm}} = \frac{\lambda_{\alpha\beta}}{M_{\rm Pl}}
(\nu_{\alpha L}\phi)(\nu'_{\beta R}\phi'),
\label{comm}
\end{equation}
where $M_{\rm Pl}= 1.2\times 10^{19}$~GeV is the Planckian mass, 
and $\alpha,\beta=e,\mu,\tau$, 
and $\phi, \phi'$ are the neutral component of the 
electroweak Higgses 
from visible and mirror sectors, 
respectively.\footnote{Here and everywhere below we 
use the Greek letters $\alpha,\beta,$... for flavor states, 
and the Latin letters $i,j,k$... for the mass states.}

After spontaneous electroweak 
symmetry breaking the Lagrangian (\ref{comm}) 
generates the terms, which mix visible and sterile neutrinos,
\begin{equation}
{{\mathcal L}_{\rm mix}}= \lambda_{\alpha\beta}
\frac{v^2}{M_{\rm Pl}}\nu_{\alpha}\nu'_{\beta},
\label{mix}
\end{equation}
where $v$= 174 GeV is VEV of the Higgses, 
which is assumed to be the same for 
$SU(2)$ and $SU(2)'$ groups. 

We assume that  the coefficients $\lambda_{\alpha\beta}$
are of the order of unity, and this is 
a natural assumption, once we consider  non-perturbative gravitational
interaction as origin of the term (\ref{comm}). In this case we have in our
model basically only one mass parameter 
\begin{equation}
\mu=v^2/M_{\rm Pl}= 5.0 \times 10^{-6}~{\rm eV} .
\label{mu}
\end{equation}
However, it is important to remark that in some models this parameter
might be slightly or essentially different:\\ 
{\em 1)} The non-perturbative gravitational 
mechanism responsible for the
communication term (\ref{comm}) could have explicit suppression
factors, flavor independent or perhaps dependent. For instance,
in the wormhole model there is an exponential suppression factor    
$\exp(-{\mathcal S})$, where ${\mathcal S}$ 
is the action, which can be approximately
expressed through the wormhole throat 
radius $R$, as ${\mathcal S} \sim M^2_{\rm Pl}R^2$.
This radius is inversely proportional to $M_{\rm Pl}$ with unknown
proportionality coefficient. The suppression can reach many orders of
magnitude \cite{linde}.\\
{\em 2)} Eq.~(\ref{comm}) can contain some  
numerical factors such as $1/(4\pi)^2$,
Clebsch-Gordan coefficients, etc.\ \\
{\em 3)} Also, one has to recall that
certain parameters of the standard model, such as gauge couplings
and top Yukawa coupling, are indeed of the order of unity, but
many other parameters are much smaller; e.g., 
$V_{ub}\approx 3\times 10^{-3}$, 
$m_s/v =7\times 10^{-4}$, and  $m_e/v =3\times 10^{-6}$.
This suggests that some flavor 
selection rule, or other mechanism, provides explicit 
suppression of Yukawa couplings. 
A similar (or the same) mechanism
could well be acting on the couplings $\lambda$, and 
result in their suppression or smallness 
(e.g.\ note that in the celebrated
seesaw model couplings similar to $\lambda$  in Eq.~(\ref{comm})
are related to the neutrino Yukawa 
couplings, that are likely to be smaller than unity).\\

The most general neutrino mass matrix in the flavor representation
can be written as 
\begin{equation}
{\mathcal L}_{\nu\ {\rm mass}} = -\frac{1}{2} (\nu,\nu')
\left( \begin{array}{cc} M & m \\ m^t & M' \end{array} \right)
\left( \begin{array}{c} \nu \\ \nu' \end{array} \right) + h.c.
\label{mdef}
\end{equation}
where $3\times 3$ matrix $m$ is given by ${{\mathcal L}_{\rm mix}}$ from 
Eq.~(\ref{mix}),
and $3\times 3$ matrixes $M$ and $M'$ are either 
complex conjugate or identical due to assumed 
mirror symmetry (see Appendix). 

We shall assume that the content of matrix $M$ in the visible sector is 
determined by interactions inside this sector, e.g.\ 
by the see-saw  mechanism. When $m=0$ the mass
matrix $M$ in the mass eigenstates basis is 
$M={\rm diag}(M_1,M_2, M_3)$, where $M_i$ are masses generated by
the see-saw mechanism. The numerical values of these masses are
outside the scope of our work.

All elements of matrix $m$ are 
approximately equal and are  mainly controlled 
by the fundamental scale of our model $\mu$. Due to $M'=M$ (or $M'=M^*$) 
the diagonalization of mass matrix (\ref{mdef}) results in maximal
mixing of $\nu_{\alpha}$ and $\nu'_{\beta}$.

Before proceeding to detailed calculations let us consider an
illustrative example of two neutrinos $\nu$ and $\nu'$. The $2\times 2$
mass matrix in this case is given by 
\begin{equation}
{\mathcal M}=
\left( \begin{array}{cc} 
M_i & \mu \\ 
\mu      & M_i 
\end{array} \right),
\label{one-nu}
\end{equation}
where we assume $M_i\gg \mu$. When the interaction between the two
sectors is switched off, $\mu=0$ and the
neutrinos are mass degenerate.
With $\mu$ taken into account, the mixing is maximal 
$\sin 2\theta=1$ and the mass eigenvalues split to 
$m_{1,2}=M_i \pm \mu$, so that  $\Delta m^2 = 4M_i\mu$
(more precisely, $\Delta m^2 = 4{\rm Re}(M_i m^*)$, since neutrino
oscillations depend on the product 
of neutrino mass matrix and its hermitian
conjugate). The transition between the split
levels results in $\nu_{\alpha} \to \nu_s$ oscillation with small 
$\Delta m^2$. 

This feature survives 
in the three neutrino case, when each mass eigenvalue, $M_i$, 
splits into two close ones (see Fig.~\ref{fig:sp}). This provides additional 
$\nu_{\alpha} \to \nu_s$ oscillation with small $\Delta m^2$ between
the split levels. We shall prove that unsuppressed oscillations 
of active to sterile neutrinos exist only between these split states of 
one level (``window'') with the small $\Delta m^2$, while the short 
wave oscillations (large $\Delta M^2$) are suppressed.

\subsection{Mass spectrum and  oscillations}
\label{m-spectrum}
In case of $m=0$ we have the 
active neutrino oscillations described by the 
matrix $M$. We introduce its decomposition 
into mass eigenstate matrix $M_i$
as:
\begin{equation}
M=U^*\ {\rm diag}(M_i)\ U^\dagger
\label{M2}
\end{equation}
The precise numerical values of masses $M_i$ and of 
the mixing angles have to be taken from the ``standard'' models of
active neutrino oscillations and 
thus they are outside the scope of our 
work.\footnote{We will consider 
the case of ``normal hierarchy'',
which--beside being consistent with all we know--arises 
most commonly in grand unified theories, models 
with flavor selection rules, etc.\ See, e.g., \cite{mu}.}
However, for two  masses 
($M_3$ and $M_2$) there are the lower bounds from
the data on atmospheric and solar neutrinos, 
\begin{equation}
\begin{array}{l}
M_3\ge (\Delta M^2_{atm})^{1/2}\approx 5\times 10^{-2}\mbox{ eV}\\[1ex]
M_2\ge (\Delta M^2_{sol})^{1/2}\approx 7\times 10^{-3}\mbox{ eV}
\label{guess}
\end{array}
\end{equation}
while the lightest mass $M_1$ remains unconstrained,
and its value can be (very) small (e.g.\ $M_1 \ll M_2$).
Also, we are allowed to borrow the mixing angles 
suggested by data, without dwelling on questions of 
how to justify their precise values. In the three neutrino case,
it is convenient to rotate away the phases in
the masses $M_i$, since they do not affect the
oscillations. We will see below that 
these phases play a more interesting role in our 6 neutrino case.
Note, that with  $M_i$ given above $\Delta m^2$ calculated for  
oscillations $\nu_{\alpha} \leftrightarrow \nu'$ ($\alpha =e, \mu,\tau$) 
in all the
three windows of Fig.~\ref{fig:sp}, respect 
the upper limits (\ref{nucl-vac}) 
and (\ref{max-mix}).

\subsubsection{Oscillations into mirror neutrinos \label{sec:oimn}}
Let us come back again to  
Eq.~(\ref{mdef}), where the matrices $M$ and $m$ are
written in the flavor representation. 
But before, we introduce a notation
for the expression of the communication term 
in the special basis where the matrixes $M=M'$ are diagonal: 
\begin{equation}
\bar{m} = U^t (m) U
\label{eq:M2}
\end{equation}

The general 6$\times$6 neutrino 
mass matrix can be brought in exact diagonal 
form using two unitary matrices, defined by:
\begin{equation}
U_{\pm}^t (M\pm m) U_\pm={\rm diag}(M_{\pm i})
\end{equation}
with $i=1,2,3$ the indices for mass eigenstates. 
The expression we get for the mixing matrix, that relates 
the ordinary and mirror neutrinos to the 6 mass eigenstates, 
$\nu^\pm_i$, is the following one:
\begin{equation}
\left\{
\begin{array}{l}
\nu=\frac{1}{\sqrt{2}} (U_+ \nu^+  - U_- \nu^-) \\
\nu'=\frac{1}{\sqrt{2}} (U_+ \nu^+ + U_- \nu^-) 
\end{array}
\label{6mix}
\right.
\end{equation}
This is the master equation,
that now we discuss and analyze.
If $m=0$, we get two equalities: 
(1) $U_\pm=U$ (namely, the two unitary matrices are equal)
and (2) $M_{\pm i}=M_i$. In this case, ordinary and mirror neutrinos
are pairwise degenerate but do not oscillate into each other as
can be verified from Eq.~(\ref{6mix}).
Instead, when $m \neq 0$, there are deviations 
from both these two equalities that, in turn, 
produce different types of oscillations. 
Their origin can be traced back to different terms of the matrix 
$\bar{m}$ of Eq.~(\ref{eq:M2}):\\
(1) The off-diagonal terms of $\bar{m}$ 
lead at first order 
to $U_+\neq U_-$ and to no-splitting between $M_{+i}$ and 
$M_{-i}$. 
Using Eq.~(\ref{6mix}), we realize that there are 
oscillations into mirror neutrinos 
connected with the splittings $M_i^2-M_j^2$, i.e.\ 
with short wavelength:
\begin{equation}
\begin{array}{l}
P_{\rm short}(\nu_\alpha\to {\rm mirror})=\frac{1}{4}\sum_{\alpha'} \\ 
\left|\sum_i (U_{+\alpha i}^* U_{+\alpha'i}
-U_{-\alpha i}^* U_{-\alpha'i})
\exp\left(-\frac{i M_i^2 L}{2 E}\right) \right|^2
\end{array}
\label{reso}
\end{equation}
(the sum over $\alpha'$ accounts for disappearance into any of the 
mirror neutrinos).
In this expression we have 
differences between $U_+$ and $U_-$, 
therefore the amplitude is linearly small 
in the parameters of $\bar{m}$, 
and this means that the oscillation probability is doubly suppressed.
For this reason, these effects are in practice negligible. \\

(2) The diagonal terms  of $\bar{m}$  
remove the double degeneracy of $M_i$-values:
\begin{equation}
M_{\pm i}=M_i \pm \bar{m}_{ii}
\end{equation}
This leads to oscillations into mirror 
states with long wavelengths, associated with the scale
\begin{equation}
\Delta m^2_i=4\, {\rm Re}(M_i\, \bar{m}_{ii}^*)
\label{eq:dmi}
\end{equation}
namely, using again Eq.~(\ref{6mix}):
\begin{equation}
P_{\rm long}(\nu_\alpha\to {\rm mirror})=\sum_{i} |U_{\alpha i}|^2\ 
\sin^2\left( \frac{\Delta m^2_i L}{4 E}\right)
\label{ave}
\end{equation}
It should be noted that, when $L/E$ 
becomes sufficiently large, this expression averages 
to 1/2: the effect is large.

Note that, using the approximation 
$U_+\approx U_-\approx U$, which has been motivated above,
and plugging in formula (\ref{6mix})
the equation $\nu_\alpha=U_{\alpha i}\nu_i$, 
one immediately obtains that 
the would-be mass eigenstates $\nu_i$ are actually 
maximal superpositions of the true mass eigenstates:
\begin{equation}
\nu_i=\frac{1}{\sqrt{2}}(\nu^+_i-\nu^-_i),
\label{summary}
\end{equation}
and similarly for $\nu'_i$:
\begin{equation}
\nu'_i=\frac{1}{\sqrt{2}}(\nu^+_i+\nu^-_i),
\label{summary1}
\end{equation}

With these equations in mind, it
is easy to summarize the pattern of oscillations: 
apart from the ordinary flavor oscillations (`short wavelength') 
we must also take into account that the state $\nu_i$ are not
mass eigenstates, but a superposition of maximally mixed 
mass eigenstates, Eq.~(\ref{summary}) 
(see again Fig.~\ref{fig:sp}). 
This leads to further oscillations
with `long wavelength', associated with the splitting 
in Eq.~(\ref{eq:dmi}).

Till now, we have presented formulae for vacuum oscillations
(Eqs.~\ref{reso} and \ref{ave}).
Occurrence of MSW effect \cite{MSW}
leads
to other cases: for instance, if  $\nu_e$
is adiabatically converted to $\nu_2$, after a 
sufficient distance, this state will further oscillate in vacuum, 
since  $\nu_2\approx (\nu^+_{2}-\nu^-_{2})/\sqrt{2}$.
At sufficiently long wavelengths, this will
produce a disappearance of 1/2 of neutrinos,
that is the same situation that was 
discussed after Eq.~(\ref{ave}).

We can summarize this subsection with the conclusion that oscillations 
to sterile neutrinos which remain unsuppressed are caused by
transitions between the split mass levels with small $\Delta m^2_i$ 
in one window shown in Fig.~\ref{fig:sp}.

\subsubsection{The scales of oscillation into mirror states \label{scales}}

The mirror model we are considering
has 3 new parameters in comparison with the usual ones:
these are the three `small' $\Delta m^2_i$ of Eq.~(\ref{eq:dmi}).
Large values of these parameters, as compared with $\Delta M^2_{sol}$,
are excluded by solar neutrino data. In this respect the most dangerous 
parameter is $\Delta m_2^2 \propto M_2$. Indeed, due to 
$U_{e3} \approx 0$ the splitting $\Delta m^2_3$ is decoupled from solar
neutrino oscillations, and $\Delta m_1^2 \propto M_1$ can always be
made very small because of the arbitrary value of $M_1$. But 
$M_2 \propto \sqrt{\Delta M^2_{sol}}$ is generically large and may 
create the problems with the second ``window'' in Fig.~\ref{fig:sp}.  
However, there exist the allowed textures of matrices $m$ ($\bar{m}$)
when oscillations in the second window are suppressed, and we
shall describe below two such examples.     

(1) Let us consider first the very specific case with  
$\bar{m} = \mu$ diag$(1,0,0)$. We shall demonstrate, in fact, that
this specific case arises from a class of initial textures of
matrix $m$ with all elements of order one, as implied by Lagrangian 
(\ref{mix}) with $\lambda_{\alpha\beta} \sim 1$. We perform rotation
of $\bar{m}$ to $m$ using  the usual mixing 
matrix $U$ with $U_{e 3}=\sin\phi= 0$, 
with the maximal atmospheric neutrino mixing angle,  
i.e.\ $\psi = 45^\circ$, 
and with a large solar angle $\omega$,
namely 
\begin{equation}
U = \left( \begin{array}{ccc}  c_{\omega} & s_{\omega} & 0 \\
-\frac {s_{\omega}}{\sqrt{2}} & \frac {c_{\omega}} {\sqrt{2}} &
 \frac {1} {\sqrt{2}} \\ 
\frac {s_{\omega}}{\sqrt{2}} & -\frac {c_{\omega}} {\sqrt{2}} &
 \frac {1} {\sqrt{2}}  
\end{array} \right), \label{eq:defU}
\end{equation}
where $s_{\omega} = \sin \omega$ and $c_{\omega} = \cos \omega$ and the 
common notation for the angles  
is $\omega=\theta_{12}$, $\phi=\theta_{13}$,  $\psi=\theta_{23}$.
Using Eq.~(\ref{eq:M2}), we obtain the communication matrix 
$m$ which has all elements ${\mathcal O}(1)$, as should be provided by
$\lambda_{\alpha\beta} \sim 1$:
\begin{equation}
m = \left( \begin{array}{ccc}
 c^{2}_{\omega} & -\frac{1}{\sqrt{8}}s_{2 \omega} & 
\frac{1}{\sqrt{8}}s_{2 \omega} \\
- \frac {1}{\sqrt{8}} s_{2 \omega} & \frac {1}{2} s^{2}_{\omega} & 
-\frac{1}{2} s^2_{\omega} \\
\frac {1}{\sqrt{8}} s_{2 \omega} & -\frac {1}{2} s^2_{\omega} & 
\frac {1} {2} s^2_{\omega}  
\end{array} \right), \label{eq:umut}
\end{equation}
This property remains true {\em generically},  
even for other values of the starting 
matrix, e.g.\ $\bar{m} = \mu$ diag$(1,i,3)$ 
($i$ here is $\sqrt{-1}$), and actually, this happens even when
$\bar{m}$ is non-diagonal. In other words, we may 
have very small or negligible oscillations 
to the second window, without violating 
the condition that the elements of $m$ are of order unity.\\
(2) As the  second example, 
we consider the case when all the elements of the communication term 
$m$ are exactly equal to one. 
That is, $m$ has the following texture:
\begin{equation}
m =  \mu \left( \begin{array}{ccc}
  1\ & 1\ & 1\ \\
  1\ & 1\ & 1\ \\
  1\ & 1\ & 1\ 
\end{array} \right). \label{eq:textm}
\end{equation}
Using Eq.~(\ref{eq:M2}) and Eq.~(\ref{eq:defU}) we get
\begin{equation}
\bar{m} =  \mu \left( \begin{array}{ccc}
  \cos^2 \omega & *  & *  \\
  {}*  & \sin^2 \omega & *  \\
  {}*  & *  & 2  
\end{array} \right). \label{eq:mbsh2}
\end{equation}
We have not written the off-diagonal 
terms of $\bar{m}$ because as explained
above they play no role in oscillations.
Note that for this texture all terms of $\bar{m}$ are 
of the same order of magnitude. In particular all diagonal 
terms are of same order 
of magnitude, i.e.\ there is no strong hierarchy 
in the diagonal terms as in the first scheme.
Therefore the splitting in the first two windows is
\begin{eqnarray}
\Delta m^2_1&\sim& 4 \mu \cos^2 \omega M_1 \cos\xi_1 \nonumber\\
\Delta m^2_2&\sim & 4 \mu \sin^2 \omega \sqrt{\Delta M^2_{sol}} \cos\xi_2
\label{dm1sch2}
\end{eqnarray}
where $\Delta M^2_{sol}$ is the solar 
neutrino mass squared difference for the large mixing angle 
(LMA) solution,
and we have introduced two phase factors $\xi_1$ and $\xi_2$.
For typical values of mixing angles in 
the LMA regime, $\sin^2\omega\sim0.3$ we get:
$\Delta m^2_1 \sim    1 \times 10^{-5}\ \mbox{eV}\ M_1$ and
$\Delta m^2_2 \sim  5 \times 10^{-8}\ \mbox{eV}^2$.
Since $M_1$ is unconstrained, we can always choose a sufficiently small
value for it, so that the splitting of the first level, or  $\Delta m^2_1$, 
becomes small and 
irrelevant for solar neutrino oscillations. 
If taken at face value, the splitting in the second
level is too large, and the net result is too much 
suppression for solar neutrinos.
However, as mentioned in Section \ref{communication} there are several
ways to escape 
this conclusion, for instance, with the help of suppression of scale $\mu$
by  $\exp(-{\cal S})$ from wormholes effects.
Or even, it is possible that $M_2$ and $\bar{m}_{22}$, when regarded 
as complex numbers, lead to a small $\Delta m^2_2$ just 
because of their relative phase (namely, the phase $\xi_2$ 
can be close to $\pi/2$). Note that in this second case, 
the smallness of $\Delta m^2_2$ can be attributed to the 
phase of $M_2$, instead than to a specific property of 
the matrix $m$.

We conclude that mirror models with gravitational communication
in many cases have a mass matrix $m$ (or $\bar{m}$) which produces 
too large oscillation effects, but there are also models compatible
with present experimental data.

\section{Applications}
In this section, we will consider the effect of oscillations into mirror
neutrinos on solar, supernova, and high energy neutrinos.
An important remark is that the atmospheric neutrino phenomenology 
is not affected by the splitting of the third level $\nu_3$, 
since for the values of the ${L}/{E}$ relevant to the
atmospheric neutrino problem, there are no oscillations into mirror
neutrinos. 
The atmospheric neutrino problem remains pure
$\nu_{\mu} \leftrightarrow \nu_{\tau}$ 
oscillations to a very good approximation, 
since CHOOZ \cite{CHOOZ} constrains the  mixing 
term $U_{e3}$ to be small.
The limit of CHOOZ implies also 
that solar neutrino oscillations are almost decoupled from 
``atmospheric'' frequency $\Delta M_{atm}^2$ in the 
usual three-flavor context.
The same remains true in our model: it is sufficient to consider 
oscillations with frequency  $\Delta M_{sol}^2$ or smaller.

\subsection{Subdominant solar neutrinos oscillations \label{subd}}
To illustrate the role of oscillations into mirror neutrinos, 
we will discuss now how the LMA solution 
of the solar neutrino problems is affected by these oscillations.
We emphasize here that our aim is not a detailed   
global fit to the solar neutrino data. Instead we want  
(1)~to demonstrate that solar neutrino data
are consistent with the presence of subdominant  oscillations 
into mirror neutrinos in our model and 
(2)~to calculate the effects produced by these oscillations,
which occur mostly at low energies.

\begin{figure*}[t]
  \begin{center}
  \mbox{\includegraphics[width=0.83\textwidth]{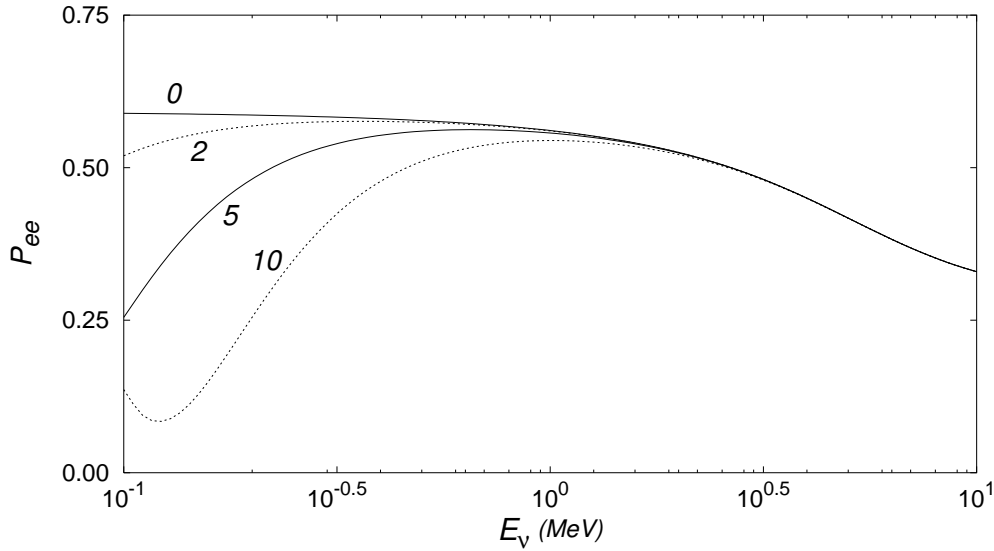}}
  \end{center}
  \caption{How the LMA survival probability 
is modified by the the oscillation into mirror states. 
The values of $\Delta m_1^2$ 
are indicated at the curves, in units of $10^{-13}$ eV$^2$.
Note the sizeable spectral distortion at low energies.
}\label{fig:pee}
  \end{figure*}

\subsubsection{CASE 1: $\Delta m^2_1\neq 0$.}

We consider here the case of sterile oscillations 
associated with the splitting of the first level. 
The splitting in the second window 
is assumed to be small as discussed 
in Sections \ref{scales} and \ref{communication}. 
The splitting of the third level 
does not affect solar neutrino oscillations.

The electron neutrino survival probability with the MSW effect 
taken into account is given by
\begin{equation}
P_{ee} =  
\cos^2 \omega\ \cos^2 \omega_m + 
          \sin^2 \omega\ \sin^2 \omega_m 
   - \cos^2 \omega\ \cos^2 \omega_m\ \sin^ 2 \delta 
 \label{eq:pee}
\end{equation}
Here, $\delta$ is the phase of vacuum oscillations
\begin{equation}
\delta =  \frac{\Delta m^2_1 \ L}{4\, E}  
\label{eq:defph}  
\end{equation}
where $L$ is the distance between production and detection,
$E$ is the neutrino energy, and $\Delta m^2_1$ is the
mass squared for oscillations into mirror neutrinos.
The mixing angle at the core of the sun 
$\omega_m$ is:
\begin{equation}
\tan 2 \omega_m  =  \frac{ \sin 2 \omega}{\cos 2 \omega - \alpha },
\mbox{ with }\alpha=\frac{2 \sqrt{2} G_F\rho_e E}{M_2^2-M_1^2}
\label{eq:tomegam} 
\end{equation}
Note that the above expression for the survival probability 
is true as long as the propagation in the 
sun is adiabatic.
This happens in the case of the LMA solution,
on which we elaborate here.

There are two distinct behaviors of $P_{ee}$ at low and high energies:\\
{\bf Low energy regime :} If the solar 
neutrino scale is the LMA mass squared difference then
at low energies $\omega_m \approx \omega$.
This implies that Eq.~(\ref{eq:pee}) becomes
\begin{equation}
P_{ee}  =   
1 - \frac {1} {2} \sin^2 2 \omega  
   - \cos^4 \omega \sin^ 2 \delta . 
\end{equation}
This can be cast in a more transparent manner:
\begin{equation}
P_{ee}  =   
      P_{ee}^{LMA} 
   - \cos^4 \omega \sin^ 2 \delta . 
 \label{eq:peele}
\end{equation}
where $P_{ee}^{LMA}$ is the standard survival probability at low
energies of the LMA solution.\\
{\bf High energy regime:} At high energies 
$\omega_m \approx \frac {\pi}
{2}$. 
This implies that Eq.~(\ref{eq:pee}) becomes
\begin{equation}
P_{ee}  = \sin^2 \omega.  
\label{eq:peehe}
\end{equation}
This is the standard survival probability at high energies of the LMA
solution.
Therefore the crucial feature of our model is that its predictions 
coincide with
the standard MSW solution at high 
energies but are affected by the subdominant
sterile oscillations at low energies: The 
standard MSW solution is modified at 
low energies, 
and most noticeably at $pp$ neutrinos energies.
This is evident from Fig.~\ref{fig:sp}, which shows  
the survival probabilities for some values 
of $\Delta m^2_1$ superimposed
on the usual LMA survival probability.

\begin{figure*}[t]
  \begin{center}
  \mbox{\includegraphics[width=0.83\textwidth]{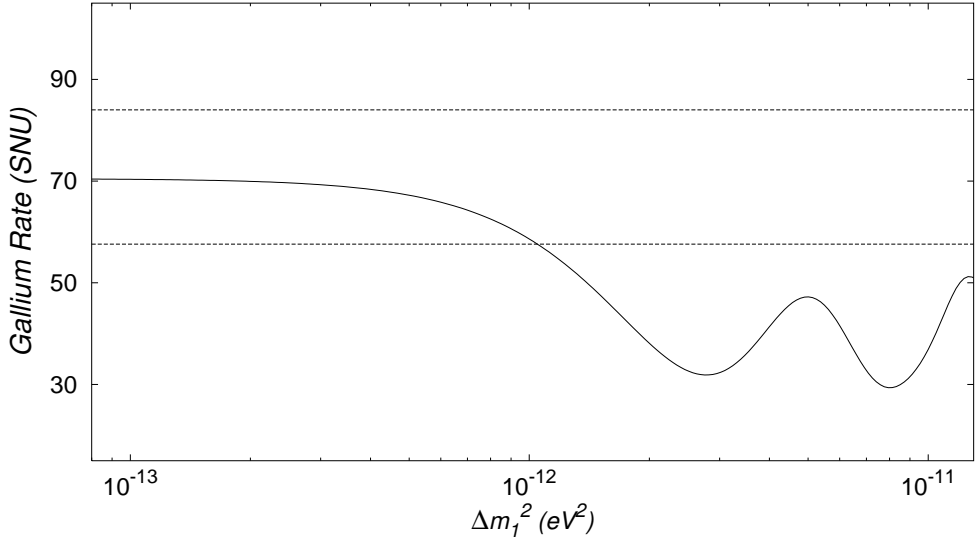}}
  \end{center}
  \caption{How the gallium rate changes with the 
new oscillation scale (the parameter $\Delta m^2_1$).
For comparison,  the 3 sigma experimental band 
is shown by dashed lines.}\label{fig:ra}
  \end{figure*}

Indeed, there is  an upper bound on $\Delta m^2_1$ 
which follows from gallium data.
For the calculation, we took the fluxes from 
\cite{bp2000} and the cross sections from \cite{bcs}, 
and used the average gallium rate
as obtained by the Gallex/GNO and SAGE experiments \cite{galli},
$70.8\pm 4.4$ SNU. 
As unperturbed case we choose the best fit LMA solution 
$M_2^2-M_1^2=6.2 \times 10^{-5}$ eV$^2$ and $\tan^2\omega=0.4$ 
as obtained in \cite{alexei}, which is consistent with the
value found in other calculations, see e.g.\ \cite{s}.
As illustrated in Fig.~\ref{fig:ra},   
the largest value  of $\Delta m^2_1$ allowed by 
the 3 sigma range is $10^{-12}$ eV$^2$.
(It is curious to note that this coincides with
the scale of the ``just-so${^2}$'' solution \cite{rag,kt,bks} 
to the solar neutrino problem.)

Two remarks are in order:\\
$(i)$ As can be seen from Fig.~\ref{fig:sp},
there is a large difference between the
profiles of the usual LMA solution and our model
in the energy range $0.2$ to $0.4$ MeV.
This is precisely the energy range of the $pp$ neutrinos. 
So future experiments like LENS \cite{lens} 
which will study $pp$ 
neutrinos in real time should see a large distortion
in the $pp$ spectrum. By contrast, the 
usual LMA solution does not predict
any energy dependent distortion in 
this energy range. When one decreases
the subdominant scale, the effect weakens though remaining still
appreciable for a reasonable range of $\Delta m^2_1$.\\
$(ii)$ The spectral 
distortion somewhat diminishes when the
mixing angle $\omega$ increases. This is 
simply due to the $\cos ^4 \omega$ factor
in front of the oscillatory term.
But even for the largest value of the angle
allowed by data, $41^\circ$, 
there is an appreciable effect for a range of $\Delta m^2_1$.

An important task is to investigate  
possible signals of seasonal variations in our scheme.
Even if the LMA solution will 
be firmly established by future experiments 
like KamLAND \cite{kamland}, 
an interesting signature of our model will be the presence of 
further seasonal variations in solar neutrino experiments,
as we now discuss.
We first rewrite our oscillation probability Eq.~(\ref{eq:pee}) in a
form suitable for studying seasonal variations.
\begin{equation}
P_{ee}  =   
      P_{ee}^{LMA} 
   - \cos^2 \omega \cos^2 \omega_{m} \sin^ 2 \delta ,
 \label{eq:peelet}
\end{equation}
where $\delta$ is defined in Eq.~(\ref{eq:defph}) and
\begin{equation}
L(t) = L_\circ \left(1 - \varepsilon \cos \Omega t \right)
\label{ld}
\end{equation}
In the above equation $\varepsilon = 1.675 \times 10^{-2}$ and
$L_\circ = 1.496 \times 10^{11}$ m. $\Omega = \frac{2 \pi} {T}$
where $T=1$ year, and $t$ is the time since the perihelion.
Hence we can write the phase of the oscillating term as
\begin{equation}
\delta = \delta_\circ \left(1 - \varepsilon \cos \Omega t \right)
\end{equation}
where in analogy with previous notation we define 
$\delta_\circ =  {\Delta m^2_1\ L_\circ}/{4 E}$. 
This modulation implies
\begin{equation}
\sin \delta = \sin \left[\delta_\circ \left(1 - 
  \varepsilon \cos \Omega t \right)
\right]          
=  \sin \delta_\circ - 
\varepsilon \delta_\circ \cos \delta_\circ \cos \Omega t
\label{eq:pht}
\end{equation}
in writing Eq.~(\ref{eq:pht}) we have used the fact that
$\delta_\circ \varepsilon \cos \Omega t \ll 1$ which is always true for
the range of $\Delta m^2_1$ we are considering, 
and for all neutrino energies
of interest.
Using Eq.~(\ref{eq:pht}) we get:
\begin{equation}
\begin{array}{rl}
P_{ee} =  & 
      P_{ee}^{LMA} 
   - \cos^2 \omega \cos^2 \omega_{m} \sin^ 2 \delta_\circ  \\
& + \varepsilon\ \delta_\circ \sin 2 \delta_\circ  
\cos \Omega t \cos^2 \omega \cos^2 \omega_m 
\equiv \left<P_{ee} \right> + P^{'}_{ee}\cos\Omega t  \label{eq:ppp}
\end{array}
\label{eq:peeft}
\end{equation}
Here, $\left<P_{ee} \right>$ is the time 
independent part of $P_{ee}$, while  
$P_{ee}^{'}$ is the amplitude of the time dependent part.

We first apply this formula to Borexino.
For an experiment which measures the monoenergetic Beryllium
line the rate, as a function of time, has the
following form
\begin{equation}
R(t)  \propto  1 + a_{\rm Be} \cos \left(\Omega t\right) \mbox{ with }
a_{\rm Be}=\frac{P_{ee}'}{\langle P_{ee}\rangle +1/5}
\end{equation}
(a factor $\approx 1/5$ comes from the neutral current contribution).
The amplitude $a_{\rm Be}$ of the time varying factor 
is a function of the
small scale $\Delta m^2_1$, and it is evaluated at the energy of the 
Beryllium line.
We find that even for the maximum value of $\Delta m^2_1$ which is
$10^{-12}\ \mbox{eV}^2$ the amplitude is only $0.001$. So there is no
significant seasonal variations at Borexino.

\begin{figure*}[t]
  \begin{center}
  \mbox{\includegraphics[width=0.58\textwidth]{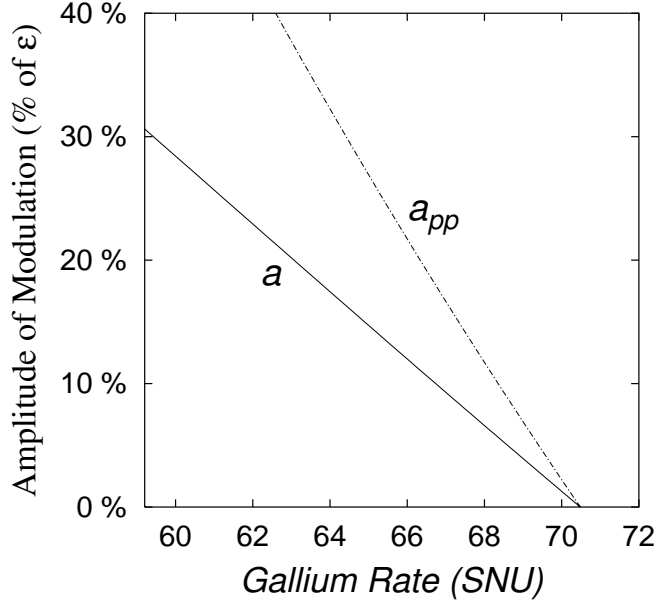}}
  \end{center}
  \caption{Amplitude $a$ of seasonal variation 
of gallium counting rate (`GNO' case) due to oscillations 
into mirror neutrinos as percentage of the excentricity $\epsilon$,
in a function of the (average) counting rate.
We also show the amplitude $a_{pp}$ of seasonal variation
for events induced by $pp$ neutrinos only (`LENS' case).
The curves are obtained varying $\Delta m^2_1$ in the range 
below $10^{-12}$ eV$^2$ for fixed LMA parameters.
}\label{fig:a}
  \end{figure*}

However, since 
the subdominant scale in our model is much smaller than the usual
``just-so'' scale, 
there may be some seasonal variations at lower energies.
With this motivation in mind, 
we analyze seasonal variations for the $pp$ neutrinos.  
We will consider two experimental situations, the first one 
{\em a la} LENS, when the $pp$ flux can be measured 
separately, the second one 
{\em a la} GNO,  when there are contributions to the signal 
also from other neutrino fluxes.

In the LENS case, the modulation of the signal is:
\begin{equation}
R(t) \propto 1 + a_{pp} \cos \left(\Omega t\right)
\end{equation}
where $a_{pp}$ is given by
\begin{equation}
a_{pp} = \frac {\int \Phi_{pp} (E) \sigma(E) P_{ee}^{'} dE} {
 \int \Phi_{pp} (E) \sigma(E) \left<P_{ee}\right> dE} .
\label{eq:app}
\end{equation}
Here $\Phi_{pp}$ is the $pp$ flux as given in BP2000 \cite{bp2000},
and $\sigma$ is the cross section for neutrino absorption on 
gallium. The integral goes from the threshold of $0.23$ MeV to the
endpoint of the $pp$ spectrum. 
For the maximum value of $\Delta m^2_1=
10^{-12} \mbox{eV}^2$ we obtain that the amplitude 
is slightly more than 
$1$ \%, which is smaller than the term $\epsilon\sim 1.7$ \% 
resulting from geometrical modulation of the distance in 
Eq.~(\ref{ld}), but is however non-negligible. 
Two conclusive remarks are in order:\\
$(i)$ It should be noted that the new modulation is in phase
with the geometrical modulation, in other terms the rate is modulated
by $(2 \epsilon +a_{pp})\cos \Omega t$.\\
$(ii)$ As mentioned above, all these 
calculations are done with a fixed 
LMA solution, namely the best fit in absence of mirror neutrino 
oscillations. If we decrease $\omega$ to the 
minimum value allowed by the LMA solution, 
the amplitude increases a bit to $1.23$ \%.

Let us consider now the case of an experiment like GNO.
When we take into account the contribution due to the other neutrino
fluxes (which do not receive seasonal variations in our model) the 
expression for $a_{pp}$ Eq.~(\ref{eq:app}) gets modified to
\begin{equation}
a = \frac {\int \Phi_{pp} (E) \sigma(E) P_{ee}^{'} dE} {
 \int \Phi_{pp} (E) \sigma(E) \left<P_{ee}\right> dE + R_{oth}}.
\label{eq:appt}
\end{equation}
where $R_{oth}$ is the contribution to the gallium experiment due to
the other neutrino sources.
So the effect will be scaled down in the actual case of GNO
(see Fig.~\ref{fig:a}). The message is that an experiment 
dedicated to measuring the $pp$ flux only will be better suited to 
look for this kind of effect.

In concluding this section we again stress the novelty of the phenomena:
Even after the MSW solution is established, the model we consider 
can lead to differences in the counting rate, spectral distortions, 
and/or seasonal variations for $pp$ neutrinos.

\subsubsection{CASE 2: $\Delta m^2_2\neq 0$.}

Let us assume that 
$\Delta m^2_1$ is strongly suppressed, but $\Delta m^2_2$ is not, 
see Sections \ref{scales} and \ref{communication}.

\begin{figure*}[t]
  \begin{center}
  \mbox{\includegraphics[width=0.83\textwidth]{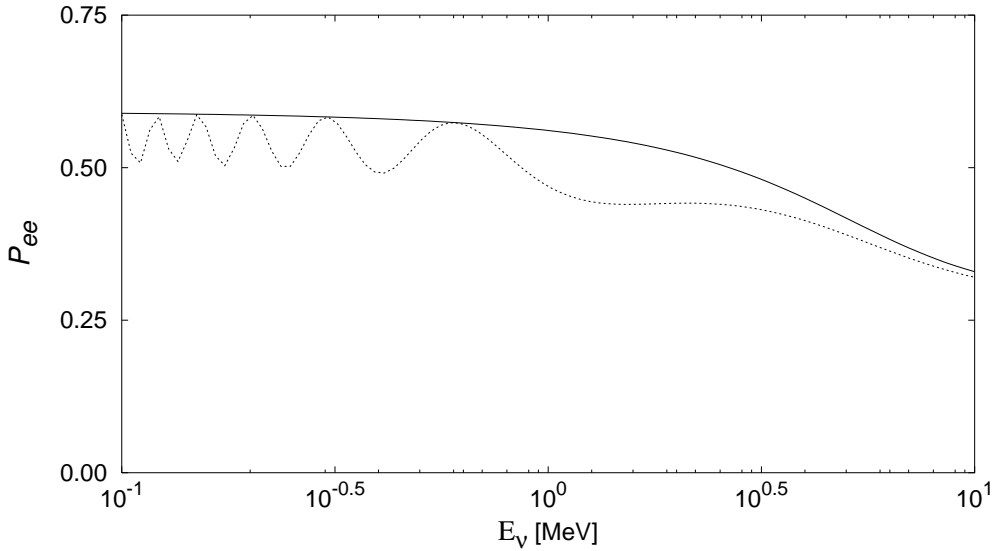}}
  \end{center}
  \caption{How the LMA survival probability (solid curve) 
is modified by the oscillation into mirror states (dotted curve,
$\Delta m^2=1 \times 10^{-11}$ eV$^2$).
}\label{fig:pee2}
  \end{figure*}

The electron neutrino survival probability becomes 
\begin{equation}
P_{ee} =   
\cos^2 \omega\ \cos^2 \omega_m + \sin^2 \omega\ \sin^2 \omega_m 
   - \sin^2 \omega\ \sin^2 \omega_m\ \sin^ 2 \delta 
 \label{eq:peesch2}
\end{equation}
(we use the notation of Eq.~(\ref{eq:defph}) for the phase of
oscillation with $\Delta m^2_1$ replaced with $\Delta m^2_2$).
In this scheme at high energies when 
$\omega_{m} \approx \frac{\pi} {2}$ we
get
\begin{equation}
P_{ee} =  \sin^2 \omega\ \cos^ 2 \delta
\label{sbf}
\end{equation}
So the usual LMA behavior is modified also at 
high energies. When we select the range 
$\Delta m^2_2<10^{-11}$ eV$^2$ there is 
no spectral distortion at boron energies, 
but just a scaling down of the usual LMA survival probability. 
This upper bound is obtained by allowing for a $10$ \%
decrease in the survival probability.
Note that the suppression of boron neutrino flux 
will be accompanied by a disappearance 
of total neutrino flux, which is given by 
the factor $\Phi=\cos^2\delta\, \Phi^0$, since only half of 
the $\nu_2$ neutrinos reaches the detector.
For this reason, the 
ratio of charged-current to neutral-current 
events yields the same $\omega$ that will be obtained by 
terrestrial experiments, say, by KamLAND. 
For the same reason, one can use the boron flux to
extract information on $\delta$ 
only if the absolute flux value is known from the
theory, and this makes the investigation difficult.

In this scheme, the gallium rate is weakly dependent 
on the new scale unlike the previous case. In fact, 
when $\Delta m^2_2$ changes from $10^{-12} \mbox{eV}^2$ to
$10^{-11} \mbox{eV}^2$ the gallium rate diminishes only 
by 3 to 7 SNU.
We plot in Fig.~\ref{fig:pee2} the modification at low 
energies of the usual LMA survival
probability due to oscillation to mirror states. 
One sees that even for the largest
value of $\Delta m^2_2$ allowed the effect is 
weak, unlike the case of the first
scheme. This is simply due to the fact that for typical 
mixing angles in the LMA region
$\sin^4 \omega \ll \cos^4 \omega$.

The analysis of seasonal variations is exactly analogous to
the analysis in the first scheme,
except that we now use Eq.~(\ref{eq:peesch2}) for the analysis.
We find that the coefficient $a$ of modulation varies 
between $\pm 25$ \% in the allowed $\Delta m^2_2$ range.
In conclusion, this scheme offers much less 
clear ``smoking-gun'' signals and its experimental 
investigation is difficult.

\subsection{Supernova neutrinos \label{ssnn}}
The framework outlined above has a natural application 
to supernova neutrinos. In fact, the 
condition that the phase of vacuum oscillation is large,
$\Delta m^2 L/4 E\gg 1$ can be rewritten as 
\begin{equation}
\Delta m^2\gg 1.3\times 10^{-19} 
\mbox{eV}^2 
\left[\frac{1\mbox{ kpc}}{L}\right]
\left[\frac{E}{20\mbox{ MeV}}\right] 
\end{equation}
where $\Delta m^2$ is the mass scale 
that gives rise to oscillations into mirror neutrinos.
We recall that 
the energy of supernova neutrino events lies certainly 
in the range $1<E<100$ MeV (lowest bound being 
mostly due to detector characteristics), 
and the distance of the 
galactic center is $L_{\rm g.c.}\approx 8$ kpc. 
We will consider two cases where mixing with 
mirror neutrinos leads to observable consequences.
The first case is based on rather standard astrophysics 
of core collapse supernovae and will be discussed at length;
the second case starts from a bolder speculation, 
about the existence of mirror supernovae, however 
it has a nicer signature. In a sense, these two cases 
correspond to the classification of oscillations into
``disappearance'' and ``appearance'' that we follow below.

\subsubsection{Disappearance of supernova neutrinos \label{ds}}
Let us consider neutrinos from a core collapse supernova.
Various patterns of oscillations into mirror neutrinos 
are possible, according to which vacuum oscillation develops 
(i.e.\ which $\Delta m^2_i$ is sufficiently large).
Here, we will focus on the 
simplest possibility: all three $\Delta m^2_i$ 
satisfy the inequality given above and therefore, 
averaged oscillations into mirror states take place
for all states. 
As shown above, see Eq.~(\ref{ave}) and following discussion,
this situation leads to the very simple
result that  half of active neutrinos of any type 
$\nu_{e,\mu,\tau}$ and their antineutrinos reach the detector.
Hence the signature of this scenario 
is that the energy observed is half the energy emitted.

The theoretical uncertainties are the
key issues to verify or contradict 
the predictions of the model we propose.
It is unclear whether the theoretical predictions of the 
energy emitted in the gravitational collapse 
${\bf E}_{\rm th}$
will become accurate enough in future 
to reveal a difference in neutrino
energy by a factor of two, say: 
${\bf E}_{\rm th}=
4\times 10^{53}\mbox{ erg}$ versus  
${\bf E}_{\rm obs}=2\times 10^{53}\mbox{ erg}$
(where we assume that oscillations into mirror 
neutrinos do take place). 
Indeed, within the existing theoretical uncertainties \cite{apr},
these two values are only marginally 
distinguishable, and can be found 
by varying the parameters of the equation of state
of the nuclear matter within the existing uncertainties.

For what regards ${\bf E}_{\rm obs}$, we recall that
a future galactic supernova 
exploding at a distance $\sim L_{\rm g.c.}$
will yield several thousands of 
neutrino events at the Super-Kamiokande detector;
similarly, very significant statistics will be 
collected at LVD, Baksan, SNO, etc. 
This should permit a determination of the total energy 
${\bf E}_{\rm obs}$ at much better than 10 \%;
possibly, without assuming energy equipartition of the 
various neutrino fluxes, but testing it by future data
(see below). 

In this respect, the $\sim 20$ neutrino events 
collected from SN1987A might seem already a statistically useful  
indication; let us look at the point more closely. 
In a recent work \cite{ll}, 
Loredo and Lamb find as optimal 
fit values ${\bf E}_{\rm obs}\sim 3\times 10^{53}$ erg
in practically all models with an accretion and a cooling 
components of the neutrino signal, 
which are expected in the 'delayed scenario' 
for supernova explosion. On examination of their 
Fig.~10, one gets convinced that oscillations into 
mirror neutrinos would not contradict present theoretical 
expectations, even though this would suggest a stiff equation 
of state for nuclear matter. 
Note incidentally that this paper, though being 
one of the the most thorough existing analyses,
assumes strict equipartition, 
does not include oscillations 
into active neutrinos, and considers only 
$\bar{\nu}_e p\to n e^+$ signal using
the `leading order' cross section. 
Equipartition is particularly 
important: for example if $\bar{\nu}_e$ carry
1/4 rather than 1/6 of the total energy, 
which is a reasonable value \cite{janka}, the best estimate 
of \cite{ll} reduces to ${\bf E}_{\rm obs}\sim 2\times 10^{53}$ erg.

\subsubsection{Appearance of supernova neutrinos \label{as}}
Mirror matter and mirror stars are expected to exist
\cite{KoOkPo,mirror,Berezh}, and this could ultimately
result into an explosion of a mirror galactic supernova.
The mirror neutrinos will oscillate into the active one,
and half of the the original flux will 
become observable.
All other radiation of mirror supernova
is undetectable, that gives the basic signature of the event:    
there will be  no optical burst    
in the direction of neutrino burst, and no radio or 
infrared radiation will 
be detected simultaneously with neutrino burst and later. 
In the case of an asymmetric gravitational collapse,
another detectable signal is the gravitational
radiation, see e.g.\ \cite{fryer}. An additional  
signature can be given by high energy mirror neutrino 
radiation from the young supernova shell \cite{vp}.

\subsubsection{Remarks on neutrino spectra and ``equipartition''}
On the top of the new effects outlined 
above, and in both cases considered in Sects.\ref{ds}
and \ref{as}, we will have also 
the (usual) effects related to flavor oscillations. 
To be specific, the 
fluxes of $\nu_e$, $\bar{\nu}_e$ and active neutrinos 
at the detector are:
\begin{eqnarray}
F_e&=& {F_{x}^0}/{2} \nonumber\\
F_{\bar e}&=& {(\cos^2\omega\, F_{\bar{e}}^0 
          +\sin^2\omega\, F_{x}^0)}/{2} \nonumber\\
F_{NC} &=& {(F_{e}^0 + F_{\bar{e}}^0 + 4 F_x^0)}/{2}.
\end{eqnarray}
where $F_{{e}}^0$, $F_{\bar{e}}^0$, 
$F_{{\mu}}^0=F_{\bar{\mu}}^0=F_{{\tau}}^0=F_{\bar{\tau}}^0\equiv F_x^0$ 
are the fluxes without oscillations. 
For definiteness, we assumed 
a normal mass hierarchy of neutrinos, 
and an angle $\phi>1^\circ$: These conditions leads to
adiabatic MSW conversion of $|\nu_e\rangle$ into $|\nu_3\rangle$ 
and of $|\bar{\nu}_e\rangle$ into $|\bar{\nu}_1\rangle$
(terms order $\phi^2\le $ few \% are neglected).
These effects--but without the oscillation into 
mirror neutrinos--have been discussed previously by a number of  
authors \cite{mvn,amol,kuo,taka,LVD}, with the generic conclusion that 
flavor mixing might lead to `hotter' ${\nu}_e$ and 
$\bar{\nu}_e$ fluxes due to the $F_x^0$ component.\footnote{This 
statement is intentionally vague, in view of the different 
conclusions reached by those who investigated SN1987A neutrinos 
including oscillations, see e.g.\ \cite{87o}, and in view of the 
fact that a full fledged theory of supernova explosions is still
missing \cite{mezza,raff,burrows,fryer}.}
The flux  of active neutrinos $F_{NC}$--from neutral current events--is 
unchanged; a peculiar signature of mirror 
oscillations is that even this flux  is reduced by a factor of 2.

Note that, by the time when  
the next galactic supernova will explode, 
we might already know the usual parameters of 
oscillations, and  therefore the usual type 
of supernova neutrino oscillations.  
In this connection, we 
believe that it is important to stress
an important consideration 
(important, even if mirror neutrino 
oscillations do not take place):
Using the large $\bar{\nu}_e$-produced data 
samples along with the 
presumably smaller samples produced by 
neutral currents and $\nu_e$, we will have 
chance to test the hypothesis of equipartition by the experimental data,
or in other words, we will be able to reconstruct the total 
energy ${\bf E}_{\rm obs}$ with a minimum theoretical bias.

Mirror neutrino oscillations  for different energy and 
situations (supernova remnant) has been recently 
discussed in Ref.\cite{c}, stressing in particular the 
possibility of a distortion of the spectrum.  
Of course a similar possibility exists for the neutrino 
spectrum from core collapse, for the 
particular case when the parameters of oscillations 
into mirror states satisfy the relation
$\Delta m^2_i L/E\sim 1$.

\subsection{High energy neutrinos \label{hen}}

High energy (HE) neutrino astronomy includes the wide range of energies
from $E \sim 100$~ GeV up to $E \sim 10^{13}$~GeV. We will consider here
the {\em diffuse} neutrino fluxes at very high energies. These fluxes can
be produced by three principal sources: accelerator sources,
topological defects (TD) and superheavy relic particles. 

There is a very 
general cascade upper limit on HE diffuse neutrino flux
\cite{cascade,book}.
   It is based on the e-m cascade which 
develops due to collisions of cascade
electrons and photons with the target  photons, e.g.\ CMB. The cascade 
is initiated by high
energy electrons or photons 
which always accompany the production of a HE neutrino.
The diffuse neutrino flux $I_{\nu}(E)$ is limited as 
\begin{equation}
E^2I_{\nu}(E) \leq \frac{c}{4\pi}\omega_{cas},
\label{cas}
\end{equation}
where $\omega_{cas}$ is the energy density of the cascade photons left
in intergalactic space. These photons have the energies in the range
of EGRET observations, which give the upper limit on the cascade
energy density $\omega_{cas} \leq 2\times 10^{-6}$~ eV/cm$^3$. 
The limit (\ref{cas}) is very general: it is valid for all processes 
of neutrino generation in extragalactic space (e.g.\ generation by TD 
and by decay of superheavy relic particles) and in the galaxies if 
they are transparent to gamma radiation.   

The only class of sources that escape the cascade upper bound (\ref{cas})
is comprised by the so called ``hidden sources'' \cite{book}. An example 
of a powerful hidden source of HE neutrinos is given by mirror matter. 
As demonstrated in Ref.\cite{BeVi} in some models the density of
topological defects in the mirror matter can be much higher than in
ordinary one. Superheavy particles produced by mirror-sector TD and 
the products of their decays are sterile in the visible world, but 
mirror neutrinos can oscillate into the visible ones. The flux of these
neutrinos can be higher than what the limit (\ref{cas}) allows.   
\subsubsection{Z-bursts}
Z-burst is a beautiful idea \cite{Zburst} of 
generation of Ultra High Energy
Cosmic Rays (UHECR) through the resonant production of Z-bosons in the
collisions of UHE neutrinos with Dark Matter (DM) neutrinos, 
$\nu+\nu_{\rm DM} \to Z^0 \to all$. The resonant energy of UHE
neutrino is $E_0=m_Z^2/2m_{\nu}=4.2\times 10^{12}m_{\rm eV}^{-1}$~GeV, 
where $m_{\rm eV}$ is the mass of DM neutrino in eV. Following 
Ref.\cite{BeVi} we calculate the number of Z-bosons produced per unit
volume and unit time, integrating over energy the diffuse neutrino flux 
$F_{\nu_i}(E)$ coupled to the Breit-Wigner cross-section $\sigma(E)$:
\begin{equation}
\dot{n}_Z=4\pi n_{\nu} \sum_i\int I_{\nu_i}(E)\sigma(E)dE=
4\pi n_{\nu}\sigma_t E_0 I_{\nu}(E_{0}),
\label{nZ}
\end{equation}
where $I_{\nu}=\sum I_{\nu_i}$, $n_{\nu}=56$~cm$^{-3}$ 
is the space density
of one flavor DM neutrinos, and 
$\sigma_t= 48\pi f_{\nu}G_F= 1.29 \times 10^{-32}$~cm$^2$ is the
effective cross-section with $G_F$ the Fermi constant and $f_{\nu}=0.019$ 
the relative width of Z-decay to neutrino channel. In Eq.~(\ref{nZ}) it is
assumed that neutrino masses are degenerate,
$m_{\nu_i}=m_{\nu}$. Using the limit on the sum of neutrino masses 
$\sum m_{\nu_i} <1.8$~eV from the spectrum fluctuations derived from 2dF
galaxy survey \cite{mlim}, we shall assume in the calculations below the 
neutrino mass $m_{\nu}=0.3$~eV as maximally allowed. 
Formula (\ref{nZ}) is exact.

In the case of decaying superheavy  particles (TD and superheavy relic
 particles) UHE photons dominate in UHECR signals   
at energy $E\geq 1\times 10^{20}$~eV \cite{BBV}. 
Their flux can be calculated as 
\begin{equation}
I_{\gamma}(E)=\frac{1}{4\pi}\dot{n}_ZR_{\gamma}(E)Q_{\gamma}(E),
\label{gamma-flux}
\end{equation}
where $R_{\gamma}(E)$ is absorption length of UHE photon and $Q_{\gamma}(E)$
is the number of photons with energy $E$ produced (via $\pi^0$ decays)
per one Z-decay. Using the observed UHECR flux at $E \sim 10^{20}$~eV,
one can calculate the flux of resonant neutrinos $I_{\nu}(E_0)$ from 
Eqs.~(\ref{nZ}) and (\ref{gamma-flux}) as 
$4\times 10^{-36}$~cm$^{-2}$s$^{-1}$sr$^{-1}$eV$^{-1}$, 
while the cascade limit
(\ref{cas}) is 5 orders of magnitude lower. 

In Ref.\cite{GeKu} it was suggested that the cascade limit can be
evaded, if X-particles decay exclusively to neutrinos, i.e.\ 
$X \to \nu\bar{\nu}$. However, in a recent work \cite{BeKaOs} it was
demonstrated that this decay results in the electroweak cascading 
in which electrons, photons and pions are efficiently produced and
thus the cascade limit (\ref{cas}) is valid for this case too. 

\subsubsection{High energy neutrinos from oscillations}

The oscillations of mirror neutrinos into the visible ones are 
characterized by oscillation length $L_{\rm osc} \sim E/\Delta m^2$,
much shorter than the typical cosmological distance $L\sim 100$~Mpc. The
only exceptional case is given by oscillation of the resonant neutrinos 
with $E_0 \approx 1\times 10^{13}$~GeV in the first ``window'' 
(see the Fig.~\ref{fig:sp}), where $\Delta m_1^2$ can be as small as 
$1\times 10^{-13}$~eV$^2$. Therefore, the average suppression due to
oscillation length is given by factor $\frac{1}{2}$. 

The conversion of the sterile neutrinos into visible ones occurs
through two stages.
Let us consider a sterile neutrino $\nu_{\alpha}'$  born with a flavor 
$\alpha$ and energy $E$. On the short length scale 
$L_{\rm short} \sim E/\Delta M^2$, where $\Delta M^2=M_i^2-M_k^2$ is the
mass squared difference of the unperturbed states, $\nu_{\alpha}'$ 
oscillates into two other sterile flavors, and we have all three 
sterile neutrinos $\nu_{\beta}'$ with $\beta=e, \mu, \tau$. 
On much longer scale $L_{\rm long} \sim E/\Delta m^2$, where $\Delta m^2$
is a scale of the window splittings, sterile neutrinos oscillate into
visible ones. Taking into account that suppression factors due to
oscillation length is 1/2, we can calculate the
probabilities $P_{\nu'\nu}$ for conversion of mirror neutrino 
$\nu'_{\alpha}$ into visible neutrino $\nu_{\beta}$, using Eqs.~(\ref{ave}) 
and (\ref{eq:defU}). In particular, for conversion of 
mirror muon neutrino $\nu'_{\mu}$ we obtain the probabilities 
\begin{equation}
P_{\nu'_{\mu}\nu_e}=\frac{\sin^22\omega}{8}, ~~~
P_{\nu'_{\mu}\nu_{\mu}}=P_{\nu'_{\mu}\nu_{\tau}}=
\frac{1}{4}-\frac{\sin^22\omega}{16},
\label{suppr}
\end{equation}
which depend only on the solar mixing angle $\omega$.  
For conversion of 
mirror tau neutrino $\nu'_{\tau}$ one should replace $\nu'_{\mu}$ by
$\nu'_{\tau}$ in Eq.~(\ref{suppr}).
For completeness we also give the relevant probabilities for
the mirror electron neutrino $\nu'_{e}$ conversion.
\begin{equation}
P_{\nu'_{e}\nu_{\mu}}= P_{\nu'_{e}\nu_{\tau}} = \frac{\sin^22\omega}{8}, ~~~
P_{\nu'_{e}\nu_{e}}=
\frac{1}{2}-\frac{\sin^22\omega}{4}.
\label{suppre}
\end{equation}
Note, that as follows from Eq.~(\ref{ave}) the probability of conversion  
$P_{\nu'_{\alpha}\nu_{\beta}}$ summed over all visible 
neutrinos $\nu_{\beta}$ is equal to $\frac{1}{2}$. 
It means that for Z-burst production 
when all neutrino flavors participate in the resonant reaction, the
total oscillation suppression $P(\nu'_{\alpha}\to \nu)=\frac{1}{2}$.

As it was already mentioned, in some cosmological models the density
of mirror-sector TD can be much larger than in the visible sector. The
ratio of neutrino fluxes from mirror and visible TD's can reach 
$2\times 10^4$ \cite{BeVi}. In our model this ratio is modified by the 
probabilities given by Eq.~(\ref{suppr}), which are different for 
different modes
of oscillations, but after summation over final states of visible
neutrinos the ratio remains the same as in Ref.\cite{BeVi}.

\section{Conclusions}

Oscillations between active and sterile neutrinos, being excluded as
the main channel of oscillations both in the solar and the 
atmospheric neutrinos,
can be interesting {\em subdominant} processes. In this paper we have
considered mirror neutrinos as the sterile ones. Our particular model 
is the ``classical'' symmetric model of the mirror matter, when apart 
from $L \leftrightarrow R$ interchange, mirror and ordinary matter
have the same interactions, coupling constants and VEV's.  
The communication between these two sectors is gravitational, it is 
described by a dimension-5 operator, suppressed by Planckian mass 
(see Eq.~(\ref{comm})). This operator can be additionally suppressed by
some factors, most strongly by $\exp(-{\cal S})$ due to wormhole transitions.
The gravitational interaction between ordinary and mirror particles,
being too weak for heavy particles of the Standard Model, can reveal 
itself in case of neutrinos. When the communication interaction is
switched off, the visible and mirror neutrinos have identical
``standard'' mass spectrum $M_i~~ (i=1,~2,~3)$, being provided e.g.\ by 
``internal'' see-saw mechanism. The communication term (\ref{comm}) 
mixes the visible and mirror neutrinos , generating a non-diagonal term 
$m$ (in the form of $3\times 3$ matrix) in the neutrino mass matrix 
(\ref{mdef}). The mixing of active and sterile neutrinos is maximal
and each unperturbed mass eigenstate with mass $M_i~~(i=1,~2,~3)$
splits into two close mass eigenstates  with masses 
$M_{\pm i}=M_i \pm \bar{m}_{ii}$ and 
$\Delta m_i^2=4{\rm Re}(M_i\bar{m}^*_{ii})$. Thus three split levels 
(windows) arise. Our basic observation is that unsuppressed
oscillations between active and sterile neutrinos occur only in 
transitions between the 
split levels of the same window. Thus, these oscillations are the
long-wavelength ones, being characterized by large 
$L_{\rm osc} \sim E/\Delta m_i^2$. These oscillations are unobservable
in atmospheric neutrinos and produce subdominant effects in solar neutrinos.

An interesting case (CASE 1, described in Section \ref{subd}) is given by
the standard LMA solution for active neutrinos with a perturbation 
term ${\mathcal L}_{\rm comm}$ (or matrix $m$) taken as in Eq.~(\ref{eq:umut}) 
of Section {\ref{scales}. In this case, only the splitting 
in the first window
works: $\Delta m_2^2$ is small and $\Delta m_3^2$ is irrelevant for solar
neutrino oscillations. In this case, the large energy effects are
practically the same as predicted by the standard LMA solution, but the flux
and the spectrum of $pp$-neutrinos are 
distorted as shown in Figs.~\ref{fig:pee}
 and \ref{fig:ra}. A rather unusual prediction is the anomalous seasonal
flux variation  shown in Fig.~\ref{fig:a}. The predicted effects,
especially distortion of the $pp$-neutrino spectrum, can be detected by
the future LENS experiment \cite{lens}. 

The mirror-visible neutrino oscillations ($\nu_{\rm mirr} \leftrightarrow
\nu_{\rm vis}$) which are 
predicted to  be observed as subdominant effects in solar
neutrinos can reveal itself as big effects 
in observation of supernova and HE neutrinos. 
In the former case (supernovae, Section \ref{ds}),
{\em half} of the expected neutrinos might actually be missed, 
though it would be crucial to improve on 
the theoretical expectation for total energy relased in neutrinos
to interpret this signal precisely; another possibility 
is finding a neutrino signal without any 
optical counterpart (Section \ref{as}). 
In the latter case  (HE neutrinos, Section \ref{hen}),
a large flux of mirror
neutrinos can be produced by mirror topological defects. Mirror neutrinos
can oscillate into the visible ones, while all accompanying mirror 
particles remain
invisible. This allows a large diffuse neutrino flux unconstrained by
e-m cascades or any other restrictions. The probability of conversion 
of mirror neutrino $\nu_{\alpha}'$ into ordinary neutrino $\nu_{\beta}$,
summed over flavors $\beta$ and averaged over distance is 
\begin{equation}
\sum_{\beta}P_{\nu_{\alpha}'\nu_{\beta}}= \frac{1}{2},
~~ \alpha,~\beta= e,~\mu,~\tau,
\end{equation}
due to the fact that oscillation length $L_{\rm osc}\sim E/\Delta m^2$
is much shorter than $L \sim 100 - 1000$~Mpc relevant for production
of high energy diffuse neutrino flux.

\section*{Acknowledgments}
We gratefully acknowledge 
discussion of the initial idea of this work and some details with Zurab 
Berezhiani. It is a pleasure to thank A.Dolgov, L.Okun, R.Raghavan,
V.Rubakov, G.Senjanovi\'c, A.Strumia and A.Vilenkin
for many fruitful discussions. 
We thank S.Moriyama for suggesting the point of mirror 
supernovae.
The work of V.B. was partially supported by INTAS (grant No.\ 99-01065).

\appendix\section{Neutrino mass matrices and transformation properties
of ordinary to mirror particles}
\label{appendix}

We shall consider here the restrictions imposed on the texture of 
neutrino mass matrices by the transformation properties of ordinary to 
mirror particles. 

Let us study the most general case of three neutrino flavors ($\alpha,\beta$)
for ordinary and mirror neutrinos $\nu_\alpha(x)$ and
$\nu_\alpha'(x)$. 
We shall limit to the case of neutrinos with ``Majorana masses''. 
For this reason, we adopt the formalism of Majorana spinors, which 
obey the relation 
\begin{equation}
\nu=C\bar{\nu}^t . 
\end{equation}
The neutrino mass part 
of the Lagrangian is given by:
\begin{equation}
{ \mathcal L}=-\frac{1}{2} \begin{array}[t]{l}
\left[M_{\alpha\beta}\ \overline{\nu_\alpha}(x) P_L \nu_\beta(x) +
M_{\alpha\beta}^*\ \overline{\nu_\alpha}(x) P_R \nu_\beta(x) +\right.\\
\left.M_{\alpha\beta}'\ \overline{\nu_\alpha'}(x) P_L \nu_\beta'(x) +
(M_{\alpha\beta}')^*\ \overline{\nu_\alpha'}(x) P_R \nu_\beta'(x) \right] -\\ 
\left.
m_{\alpha\beta}\ \overline{\nu_\alpha}(x) P_L \nu_\beta'(x) -
m_{\alpha\beta}^*\ \overline{\nu_\alpha}(x) P_R \nu_\beta'(x) \right.
\label{lagg}
\end{array}
\end{equation}
where the star  denotes complex conjugation and $P_{L,R}$ are the projection 
operators.
It is easy to see that this is the most general case, 
by recalling that for the Majorana spinors 
$\bar{\lambda}P_{L,R} \chi=\bar{\chi} P_{L,R}\lambda$. 
The parameters of this Lagrangian 
(apart from the arbitrary phases of the fields) 
are those of the two symmetric matrices $M$ and $M'$, and 
of the arbitrary matrix $m$, and we can construct from them 
the neutrino mass matrix:
\begin{equation}
{\mathcal M}=\left( \begin{array}{cc} M & m \\ m^t & M' \end{array} \right)
\label{general}
\end{equation}

In the context of quantum field theory, the 
mirror symmetry implies that the action ${\mathcal S}=\int d^4 x {\mathcal L}$ 
is invariant under left$\leftrightarrow$right transformation of the
fields (see Lagrangian (\ref{lagg}) as a particular example).  

Two kinds of left$\leftrightarrow$right 
transformations of the arbitrary fermionic fields $\Psi$ and $\Psi'$,
generically denoted as
\begin{equation}
\Psi_{L,R} \leftrightarrow \Psi_{R,L}'
\label{generic}
\end{equation} 
can provide this invariance.

(1) The first one is given by 
\begin{equation}
\Psi(x)\leftrightarrow C\ \overline{\Psi'}^t(x).
\end{equation}
It is easy to convince oneself that this transformation 
belongs  to the class described by Eq.~(\ref{generic}): e.g.
the ordinary left current is transformed into a  
mirror right-current.
Applying this transformation to  the neutrino mass 
Lagrangian (\ref{lagg}), 
we obtain that the action is invariant if the following conditions hold:
$M_{\alpha\beta}=M'_{\alpha\beta}$, 
$m_{\alpha\beta}=m_{\beta\alpha}$.
Then, the neutrino mass matrix reads:
\begin{equation}
{\mathcal M}=\left( \begin{array}{cc} M & m \\ m & M \end{array} \right)
\label{type1}
\end{equation}
with the conditions that $M$ and $m$ are symmetric.

(2) Now let us consider the standard case 
of mirror transformation 
\begin{equation} 
\Psi(t,\vec{x})\leftrightarrow  \gamma^0 \ \Psi'(t,-\vec{x}).
\end{equation} 
Again, this transformation converts  left- into right-currents,
and belongs to the class described by  Eq.~(\ref{generic}).
Applying this to  the Lagrangian given by Eq.~(\ref{lagg}), 
and noting that the measure of integration $d^4x$ is invariant under 
inversion, 
we obtain that the action is symmetric if the following conditions hold:
$M_{\alpha\beta}=(M'_{\alpha\beta})^*$ and 
$m_{\alpha\beta}=(m_{\beta\alpha})^*$.
Then, the neutrino mass matrix reads:
\begin{equation}
{\mathcal M}=\left( \begin{array}{cc} M & m \\ m^t & M^* \end{array} \right)
\label{type2}
\end{equation}
with the conditions that $M$ is symmetric and $m$ is hermitian.

In most applications considered in this paper 
one can take $M=M^*$ and $m=m^*$, and thus the 
cases (1) and (2) are identical. 
However, there can be the cases
when the two mass matrices 
in Eqs.~(\ref{type1}) and ({\ref{type2}) lead to  
different physical situations. 
Consider, for example, the one family case, when the element of the
matrix $m$ is a small value of
order $\epsilon$ and the element of $M$ is order 1 (that is the 
case we have in mind in our study). Let us further
assume that the first parameter is real and the second
is a pure phase factor $e^{i\xi}$
(we use a very similar notation in Eq.~(\ref{dm1sch2})). 
In the first and second case, we have respectively:
\begin{equation}
{\mathcal M}_1
=\left( \begin{array}{cc} e^{i\xi} & \epsilon \\ 
  \epsilon & e^{i\xi} \end{array} \right)\ \ , \ \ 
{\mathcal M}_2
=\left( \begin{array}{cc} e^{i\xi} & \epsilon \\ 
  \epsilon & e^{-i\xi} \end{array} \right)
\end{equation}
Oscillations are described by the 
combination ${\mathcal M} {\mathcal M}^\dagger$. Neglecting the  terms of
order of $\epsilon^2$, we obtain:
\begin{equation}
{\mathcal M}_1{\mathcal M}_1^\dagger
=\left( \begin{array}{cc} 1 & 2 \epsilon\; \cos\xi \\ 
2 \epsilon\; \cos\xi & 1 \end{array} \right) , 
\end{equation} 
in case (1) and 
\begin{equation}
{\mathcal M}_2{\mathcal M}_2^\dagger
=\left( \begin{array}{cc} 1 & 2 \epsilon\; e^{i\xi} \\ 
2 \epsilon\; e^{-i\xi} & 1 \end{array} \right),
\end{equation}  
in case (2).
The mixing angle is in both cases maximal, but in the first 
case  $\Delta m^2$ can be much smaller than $\epsilon$ 
(when the phase $\xi$ is close to $\pm \pi/2$), while in the second
case it is always of order of $\epsilon$.

%\newpage
\footnotesize 
\frenchspacing
\begin{multicols}{2}

\end{multicols}
\end{document}